# Quantification of dipolar interactions in $Fe_{3-x}O_4$ nanoparticles

Carlos Moya,[*,†] Óscar Iglesias,[†] Xavier Batlle,[†] Amílcar Labarta[†]

[†] Departament de Física Fonamental and Institut de Nanociència i Nanotecnologia (IN2UB), Universitat de Barcelona, Martí i Franquès 1, 08028 Barcelona, Spain

KEYWORDS (magnetite nanoparticles, silica coating, dipolar interactions, numerical simulation)

**ABSTRACT:** A general method for the quantification of dipolar interactions in assemblies of nanoparticles has been developed from a model sample constituted by magnetite nanoparticles of 5 nm in diameter, in powder form with oleic acid as a surfactant so that the particles were solely separated from each other through an organic layer of about 1 nm in thickness. This quantification is based on the comparison of the distribution of energy barriers for magnetization reversal obtained from time-dependent relaxation measurements starting from either (i) an almost random orientation of the particles' magnetizations or (ii) a collinear arrangement of them prepared by previously field cooling the sample. Experimental results and numerical simulations show that the mean dipolar field acting on each single particle is significantly reduced when particles' magnetizations are collinearly aligned. Besides, the intrinsic distribution of the energy barriers of anisotropy for the non-interacting case was evaluated from a reference sample where the same magnetic particles were individually coated with a thick silica shell in order to make dipolar interactions negligible. Interestingly, the results of the numerical simulations account for the relative energy shift of the experimental energy barrier distributions corresponding to the interacting and non-interacting cases, thus supporting the validity of the proposed method for the quantification of dipolar interactions.

## Introduction

Currently, systems formed by magnetic nanoparticles (NPs) are widely investigated due to their potential applications in both information storage and biomedicine, such as for example in hyperthermia, as contrast agents for magnetic resonance imaging and for drug delivery.[1-4] In particular, magnetite ($Fe_{3-x}O_4$) NPs are among the most commonly studied materials due to their low-toxicity, easy production and functionalization, and good magnetic performance, as they exhibit a relatively high specific magnetization and an intermediate value of the magnetocrystalline anisotropy.[5-7] However, the latter could be also considered as an important drawback for biomedical applications of colloidal suspensions of NPs, since it facilitates the formation of big particle aggregates through dipolar interactions among them, either provoking blood thrombi or substantially modifying their magnetic response.[5-8] For instance, recent works have shown a significant variation on the heating efficiency for hyperthermia of a colloidal suspension of NPs depending on the particle concentration and their aggregation state in the heating medium.[9-11] Therefore, proposing a simple method for the quantitative characterization of the inter-particle interactions could be very useful to improve their magnetic performance in such applications. Experimentally, there are two strategies to study the effect of inter-particle interactions in a nanoparticle system. The first one consists on the study, as a function of the particle concentration, of either a colloidal suspension of NPs or a solid matrix where NPs are embedded.[12-13] However, the principal drawback of this method relies on the wide distribution of inter-particle distances and the consequent little accuracy in the determination of a representative mean value, which complicates the interpretation of the data and its generalization. The second strategy is based on the coating of the NPs with either a polymer or an inorganic corona to produce a kind of individual core-shell nanostructure, which in a powder form enables the magnetic cores to keep a quite regular distance from each other. This is indeed a more suitable method than the former to perform a quantitative determination of the inter-particle interactions.[14] In particular, silica coating is an ideal alternative in which particles can be isolated and separated from each other over long distances, so that dipolar interactions can be tuned as a function of the inter-core distance and eventually drastically reduced.[15] On the other hand, numerical simulations have proven to be a useful tool to investigate both the magnetic properties of single NPs and the collective behavior of interacting ensembles of them, providing a general methodology to understand and support experimental results.[10,16,17]

In this framework, we have quantified the effect of dipolar interactions on the magnetic properties of 5 nm $Fe_{3-x}O_4$ NPs in powder form with oleic acid as a surfactant so that the latter separates the NPs from each other solely through an organic layer of about 1 nm in thickness. We propose a simple method to detect the presence of dipolar interactions that consists on the comparison of the distributions of energy barriers for magnetization reversal obtained from time-dependent relaxation measurements, starting from configurations with either random orientation of the particles' magnetizations or collinear arrangement of them prepared by previously field cooling the sample. The validity of these results is supported by both macrospin simulations of a poly dispersed ensemble of spherical NPs and the intrinsic distribution of the energy barriers of anisotropy of the NPs calculated from a reference sample where NPs are individually coated by a thick silica shell in order to make dipolar interactions negligible.

**Preparation of the samples and experimental techniques**

Two samples have been thoroughly studied and compared. Sample R1 was synthesized by thermal decomposition of the organo-metallic precursor Fe(acac)$_3$ in the presence of oleic acid and oleylamine as surfactants and 1,2-hexadecanediol as stabilizer agent, following a method extensively described elsewhere.[7-18] Monodisperse magnetite Fe$_{3-x}$O$_4$ NPs of about 5 nm in diameter were obtained. In order to make dipolar interactions among the particles negligible keeping them far apart from each other, magnetite NPs in sample R1 were individually coated with SiO$_2$ by using a microemulsion method based on a solution of tetraethoxysilane (TEOS) in water-in-oil (w/o), since this is a reproducible procedure where the thickness of the SiO$_2$ layer and number of magnetite cores inside each silica shell can be easily controlled through the fraction of TEOS and the concentration of Fe$_{3-x}$O$_4$ NPs.[19,20] This sample is called R2 hereafter.

Samples were prepared for transmission electron microscopy (MT80-Hitachi microscope) by placing one drop of a dilute suspension of NPs onto a carbon-coated cooper grid and letting it dry at room temperature. The size distribution was analyzed by measuring at least 2000 particles and the resultant histograms are shown in Fig. S1a and S1b, Supporting information. Both the crystalline quality of individual NPs and the core-shell nanostructure of the silica-coated NPs were characterized by high-resolution TEM (HR-TEM) micrographs obtained by Titan high-base microscopy (see Fig. 1b-d).

The crystal phase of the iron oxide particles was identified by powder X-ray diffraction performed in a PANalytical X'Pert PRO MPD diffractometer by using Cu Kα radiation (see Fig. S2, Supporting information). The patterns were collected within 5 and 120° for 2θ. The XRD spectra were indexed to an inverse spinel structure.

The Fe content in the samples was determined by inductively coupled plasma-optical emission spectrometry (ICP-OES) by using a Perkin Elmer model OPTIMA 3200RL after digesting the samples in a mixture of HClO$_4$:HNO$_3$ 5:25, and diluting them with distilled water.

Magnetization measurements were performed with a Quantum Design SQUID magnetometer. Hysteresis loops $M(H)$ were measured at several temperatures within 2 and 300 K to study the saturation magnetization ($M_s$) and the coercive field ($H_c$) under a maximum magnetic field of ± 20 kOe. $M_s$ was obtained by extrapolating the high-field region of $M(H)$ to zero field, assuming the high-field behavior $M(H) = M_s + \chi \cdot H$, where $\chi$ is a residual susceptibility. $M_s$ values were normalized to the magnetic content evaluated from ICP-OES measurements.

The $M_s$ values obtained from the hysteresis loops at 2 K ($M_s = 82 \pm 2$ emu/g and $M_s = 76 \pm 2$ emu/g for R1 and R2, respectively) were just slightly smaller than the bulk one,[21] indicating an almost perfect ferrimagnetic order throughout the whole NPs, as previously observed in samples with very high crystalline quality synthesized by the thermal decomposition method.[7] The coercive field was calculated as $H_c = (|H_c^+| + |H_c^-|)/2$ and the obtained values at 2 K were 690 ± 30 Oe and 650 ± 20 Oe for R1 and R2, respectively.[7,22]

The thermal dependence of the magnetization was studied after zero field cooling ($M_{ZFC}$) and field cooling ($M_{FC}$) the samples. These curves were measured using the following protocol: the sample was cooled down from 300 to 2 K in zero magnetic field, then a static magnetic field of 50 Oe was applied and $M_{ZFC}$ was measured while warming up from 2 to 300 K; once room temperature was reached, the sample was cooled down again to 2 K while 50 Oe was applied; after that the sample was warmed up to 300 K and $M_{FC}$ was collected under the applied field of 50 Oe.

The time dependence of the thermo-remnant magnetization in the temperature range 2-30 K was measured after field cooling sample R1 under various fields (50, 200 and 1000 Oe) and sample R2 under 50 Oe, from room temperature down to the measuring temperature, at which the magnetic field was switched off. The magnetization decay was then recorded as a function of time at zero field and at several temperatures.

**Structural characterization**

TEM images of R1 show spherically shaped particles with a narrow size distribution (Fig. 1a) that was fitted to a log-normal function with values of the mean diameter $D_{TEM} = 5.3$ nm and the unitless standard deviation of σ= 0.20 (Fig. S1a, Supporting information). Fig. 1a also shows the regularity and monodispersity of the NPs. Besides, HR-TEM image in Fig. 1b demonstrates the high crystalline quality of individual NPs. The mean particle diameter estimated from X-ray data ($D_{XRD} = 5.1 \pm 0.2$ nm) is in agreement with $D_{TEM}$ supporting also the absence of crystalline defects and the fact that the NPs are single-crystal domains. Silica coating of those NPs produced highly regular core-shell nanostructures (sample R2) containing a single magnetic core in most of the cases (see Fig. 1c). The values of the mean diameter and the unitless standard deviation obtained from the histogram of the size distribution are $D_{TEM} = 44$ nm and σ= 0.19, respectively. HR-TEM image in Fig. 1d indicates that the morphology of the magnetite NPs remains unchanged after silica coating. The thickness of the silica layer coating the magnetite cores is about 20 nm, which is more than enough to make inter-particle dipolar interactions negligible.

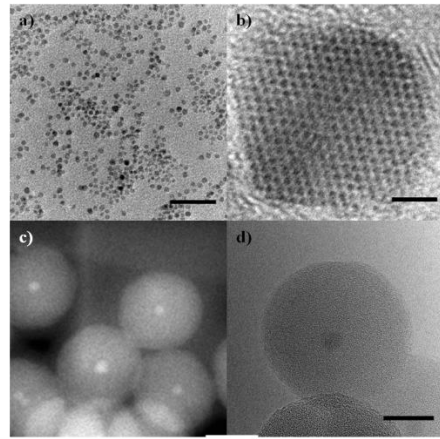

**Figure 1.** TEM characterization of samples R1 and R2: a) Low-TEM resolution image for R1 NPs. b) High-resolution TEM image for R1. c) Low resolution image of silica-coated NPs (sample R2). d) HAADF image of several silica-coated NPs. Scale bars: a) 50 nm, b) 1 nm c) 30 nm and d) 15 nm.

## Magnetic characterization

To get a basic idea of the effect of dipolar interactions on the magnetic properties of the NPs, isothermal magnetization curves for samples R1 and R2 were measured as a function of the magnetic field up to 20 kOe for selected temperatures within 150 K and 300 K, where NPs of the two samples are fully superparamagnetic (SPM). The obtained magnetization curves $M$ as a function of the scaling variable $H/T$ are plotted in Fig. 2. The excellent superposition of the scaled magnetization curves for sample R2 (see main panel of Fig. 2) is a clear indication of its SPM behavior and corroborates the nonexistence of dipolar interactions among the particles when the magnetic cores are well-separated from each other by the silica coating. However, this kind of scaling is not achieved in sample R1 (see inset to Fig. 2), where particles are solely coated by oleic acid and dipolar interactions are significantly affecting their magnetic response.

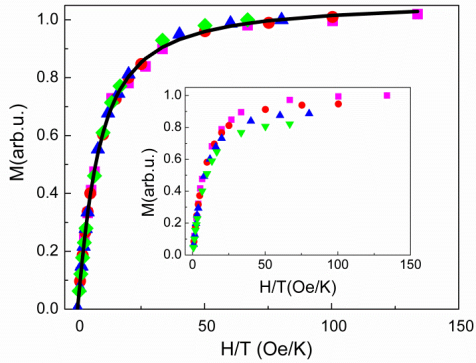

**Figure 2.** Isothermal magnetization curves for sample R2 measured at 150 K (lilac square), 200 K (red circles), 250 K (blue triangles) and 300 K (green diamonds), plotted as a function of $H/T$. Inset: Isothermal magnetization curves for sample R1 plotted as a function of $H/T$. The continuous line is the fit to Eq. (1).

These qualitative results are also confirmed by comparing $M_{ZFC}$ - $M_{FC}$ curves for the two samples (see Figure 3). Whereas $M_{ZFC}$ - $M_{FC}$ data for R2 exhibit the common trends corresponding to non-interacting SPM particles undergoing a blocking process - a sharp peak in $M_{ZFC}$ and a monotonous increase in $M_{FC}$ as the temperature is lowered -, the enhancement of dipolar interactions in sample R1 as compared to sample R2 is suggested by (i) the blocking temperature indicated by the position of the peak in $M_{ZFC}$ is multiplied by a factor of about 1.3, (ii) the ZFC peak is significantly broader, and (iii) $M_{FC}$ becomes flattened below the blocking temperature.[12] Interestingly, all those differences between the $M_{ZFC}$-$M_{FC}$ curves for the two samples may entirely be attributed to the effect of the inter-particle interactions since the magnetic cores are exactly the same. We can take advantage of this fact in order to get a first quantitative estimation of the changes caused by dipolar interactions on the effective distribution of energy barriers of the NPs.

We first fitted the scaled magnetization curves of sample R2 to

$$M(H,T) = M_S \frac{\int m P(m) L(mH/k_B T) dm}{\int m P(m) dm} + \chi_p H \qquad (1)$$

where we assume that, in the SPM regime and for non-interacting particles, the magnetization can be described by averaging the Langevin function $L(x)$ accounting for the magnetization $m$ of each particle with the log-normal distribution $P(m)$ of the magnetic moment of the particles[23] plus a linear-field contribution originating at a residual paramagnetic susceptibility $\chi_p$. Taking into account that $m = M_S V_m$, where $V_m$ is the activation magnetic volume of each particle, the fitted $P(m)$ distribution can be transformed in the distribution of activation magnetic volumes being the obtained values of the unitless standard deviation and the mean magnetic diameter $\sigma_V = 0.63$ and $D_m = 5.3$ nm, respectively, in good agreement with the structural parameters deduced from TEM images. This concordance between the structural and magnetic-volume parameters also supports both the high crystalline quality of the NPs and the nonexistence of significant inter-particle interactions for sample R2.

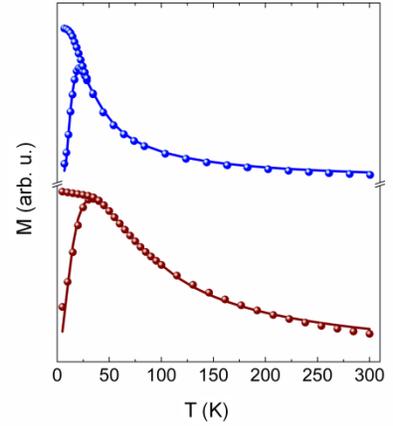

**Figure 3.** ZFC-FC magnetizations (H=50 Oe). Brown and blue spheres correspond to samples R1 and R2, respectively. Solid blue line and solid brown line correspond to the fit of the ZFC curves for R1 and R2, respectively, to Eq. (2).

It is well-known that the ZFC susceptibility is more sensitive to the distribution of particle volumes than the equilibrium magnetization curves.[24] Consequently, $\sigma_V$ can be further refined by performing a fit of $M_{ZFC}/H$ to the following expression[25-27]

$$\frac{M_{ZFC}(T)}{H} = \frac{1}{3k_B T} \int_0^{M_S V_p(T)} m^2 P(m) dm + \frac{M_S}{3K} \int_{M_S V_p(T)}^{\infty} m P(m) dm \qquad (2)$$

deduced from the Gittleman's model,[25] where $D_m$= 5.3 nm was fixed to the value obtained from the fitting of Eq. (1) to the magnetization curves. In Eq. (2), the first term accounts for the contribution of SPM particles [with volumes $V<$ $V_p(T) = \frac{25k_B T}{K}$] and the second is for the blocked ones [$V > V_p(T)$]. Here $k_B$ is the Boltzmann constant and $K$ the effective anisotropy. As a result of the fitting procedure, we obtained $\sigma_V = 0.54$ and $K = 3.1 \times 10^5$ erg/cm$^3$, so that the effective anisotropy is about three times the corresponding one for bulk magnetite ($K = 1.1 \times 10^5$ erg/cm$^3$),[22] as expected for NPs of a few nanometers in diameter where surface anisotropy is the dominant contribution.[28] Applying the same kind of fitting to the ZFC susceptibility of sample R1 and imposing $D_m = 5.3$ nm we obtained $K = 3.7 \times 10^5$ erg/cm$^3$ and $\sigma_V = 0.81$ which provides a qualitative estimation of the effect of the inter-particle interactions on the distribution of activation magnetic volumes. From these results we can conclude that inter-particle interactions slightly increase the effective anisotropy of the particles but they significantly broaden the distribution of activation magnetic volumes, broadening as well the distribution of effective energy barriers.

The existence of high energy barriers originating from inter-particle interactions has also been studied by measuring the time relaxation of the magnetization at several temperatures following the protocol detailed in Sec. 2. The obtained relaxation curves were analyzed within the context of the so-called $T\ln(t/\tau_0)$ scaling where,[29] in order to make all the curves collapse onto a single master curve, the characteristic attempt time was set to $\tau_0 = (5 \pm 4) \times 10^{-10}$ s. The results of this scaling for R1 and R2 after field cooling the samples under the relatively low field of 50 Oe are shown in Fig. 4. The most prominent difference between these two relaxation curves corresponding to the interacting and non-interacting cases, is the extra magnetic viscosity slowing down the magnetic relaxation in sample R1 as a consequence of the magnetic frustration introduced by inter-particle interactions, which are almost completely suppressed in sample R2 by the silica coating.

Besides, it is also evident that the magnitude of the cooling field applied before the relaxation is measured dramatically affects the time decay of the magnetization for sample R1, lowering the magnetic viscosity and making the relaxation look quite similar to the non-interacting case (R2) at intermediate values of the cooling field (200 Oe). As shown in Sec. 5, the latter is a direct consequence of the reduction of the average value of the dipolar field acting on each particle as the initial configuration of the particles' magnetizations becomes collinear through the effect of increasing the cooling field.

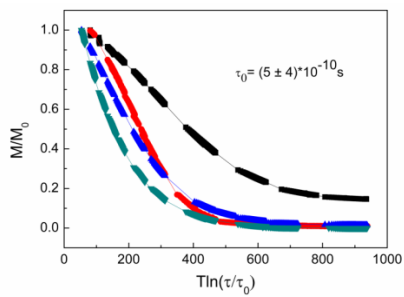

**Figure 4.** Scaling of the relaxation curves measured at several temperatures (2-30K) with an attempt time of $\tau_0 = (5 \pm 4) \times 10^{-10}$ s, after field cooling at a field of 50 Oe for R1 (black line) and R2 (red line), and 200 Oe (blue line) and 1000 Oe (green line) for R1.

Consequently, we propose the study of the relaxation master curve as a function of the cooling field as a simple method to gain a rough estimation of the effect of dipolar interactions on the effective energy landscape of a nanoparticle system. Of course, this method does not provide an exact evaluation of the contribution of the dipolar interactions to the magnetic energy of the system because of both the incomplete suppression of dipolar interactions and the modification of the energy barriers of anisotropy by the cooling field, but at least it can be taken as a clear proof of the existence of inter-particle interactions and a means to get a first approximation of the energy involved.

On the other hand and as shown previously,[30] the $T\ln(t/\tau_0)$ scaling allows determining the effective distribution of energy barriers $f(E)$ explored by the system along the relaxation process by performing the numerical derivative of the master curve with respect to the scaling variable. The distributions so obtained for samples R1 and R2 are shown in Fig. 5. It is worth noting that $f(E)$ for the sample with inter-particle interactions (R1) extends to higher energies and that the energy of the maximum is also higher than for R2 in agreement with the effective distributions of energy barriers obtained by fitting the ZFC data. Thus, the peak of the energy distribution for R2 is placed at a temperature $T_{max}$= 171 K that is comparable to that corresponding to the mean anisotropy energy barrier given by $KV_m/k_B$= 175 K, confirming that in sample R2 dipolar interactions are negligible. The peak for sample R1 is shifted to $T_{max}$= 217 K presumably due to the influence of dipolar interactions on the original distribution of energy barriers, and, as will be shown in Sec. 5, this can be simply accounted through the changes introduced by the local dipolar fields on the anisotropy barriers.

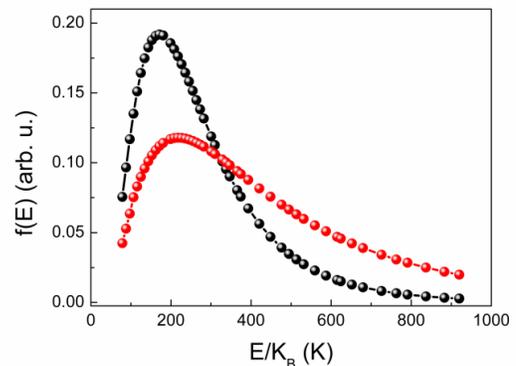

**Figure 5.** Effective distribution of energy barriers obtained from the scaling curves in Fig. 4 for R1 (red spheres) and R2 (black spheres).

## Numerical simulation

In order to gain a deeper insight on the origin of the dependence of the relaxation behavior and the associated effective distribution of energy barriers observed experimentally as a function of the cooling field in a nanoparticle system with inter-particle interactions, we have conducted a series of numerical simulations to compute the dipolar fields and energy barriers of a model of nanoparticle assemblies which mimics the experimental ones. For this purpose, nanoparticle assem-blies have been built by placing a fixed number $N$ of spheres with diameter $D$ [assemblies with a lognormal distribution of particle sizes $f(D)$ have also been considered] with different spatial arrangements inside a cubic box of varying size $aL$ (where $a$ is the lattice spacing, taken as 1 nm hereafter), adjusted so that the desired volume concentration defined as $c = \frac{V_N}{(aL)^3}$ is achieved ($V_N$ is the total volume of the nanoparticle). We consider that spheres do not overlap and we impose a minimum inter-particle separation given by the thickness of the surfactant layer. We work in the macrospin approximation so that the total particle magnetization is given by $\vec{m}_i = M_s V_i \vec{S}_i$ where $\vec{S}_i$ is a Heisenberg 3D spin of modulus 1, $M_s$ is the saturation magnetization and $V_i$ the particle volume. The interaction energy considered has contributions from magnetic anisotropy, dipolar interaction and Zeeman coupling with the field:

$$E_{ani} = -\sum_i K_i V_i (\vec{S}_i \cdot \vec{n}_i)^2,$$
$$E_{dip} = -M_s \sum_i V_i (\vec{S}_i \cdot \vec{H}_i^{dip}), \quad (3)$$
$$E_H = -\mu_0 M_s \sum_i V_i (\vec{S}_i \cdot \vec{H})$$

where $\vec{n}_i$ are the uniaxial anisotropy directions, $H$ the magnetic field, and we have defined the dipolar field acting in the ith particle as:

$$\vec{H}_i^{dip} = -\frac{\mu_0}{4\pi a^3} M_s \sum_{\{j\}} V_j \left[ \frac{\vec{S}_j}{r_{ij}^3} - 3\frac{(\vec{S}_j \cdot \vec{r}_{ij})\vec{r}_{ij}}{r_{ij}^5} \right] \quad (4)$$

being $r_{ij}$ the interparticle distances in units of $a$. Typical energy scales for the three contributions for the case of sample R1 can be obtained by setting $M_s = 5 \times 10^5$ A/m and $K = 3.1 \times 10^4$ J/m$^3$. In particular, considering a mean diameter of $D_m = 5.3$ nm and an inter-particle separation $r_{ij} = d = 7a$, which corresponds to a thickness of the surfactant layer of about $a$ (1 nm in the case of sample R1) and a volume concentration of $c = 0.3$ assuming a close packing of the particles, we get $E_{dip} = \frac{\mu_0}{4\pi d^3}(M_s V_m)^2 = 4.43 \times 10^{-22}$ J $= 32.1$ K for the characteristic dipolar energy in sample R1. Typical dipolar fields in this sample are of the order of $H_{dip} = \frac{\mu_0}{4\pi d^3} M_s V_m = 114$ Oe, to be compared with the mean anisotropy field of $H_{ani} = \frac{2K}{M_s} = 1240$ Oe.

We have first computed the local dipolar fields generated by $N = 4096$ particles randomly distributed in a box of size $aL$ at various concentrations $c$. Since dipolar fields depend on the magnetic configuration of the assembly, we have considered two extreme cases: magnetic moments i) aligned along z axis and ii) randomly oriented. In Fig. 6, we show the histograms of the modulus of the dipolar field for the two mentioned magnetic configurations and three values of $c$. Only the local dipolar fields generated at the sites of the 1000 particles at the central part of the box have been taken into account in order to minimize finite-size effects. Notice that the spread in dipolar field values around a mean value that depends on particle concentration $c$ is caused by the random position of the particles in the simulation box. The exact distribution of $H_{dip}$ values depends also on the magnetic configuration. For increasing concentrations, the spread of $H_{mod}$ histograms increases for the random case while it remains more or less constant for the aligned particles. Most importantly, we observe that, for all concentrations, the magnitude of the dipolar fields is clearly reduced when aligning the particles' magnetization along the z axis. It is worth noting that the situation for sample R1 corresponds approximately to the case with $c = 0.3$, since this is the value of the volume concentration that can be estimated for this sample.

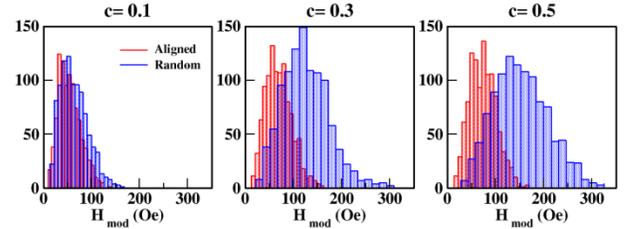

**Figure 6.** Histograms of the dipolar field moduli for 4096 particles randomly placed in a cubic box of size $aL$ for three different concentrations c = 0.1, 0.3, 0.5, from left to right. Histograms in red are for a configuration with magnetic moments aligned along the z axis while blue ones correspond to randomly oriented magnetic moments. Only contributions from 1000 particles in the central part of the considered box have been taken into account.

This can be more clearly seen in Fig. 7, where the dependence of the mean value of the modulus of the dipolar field on the concentration is plotted. Even for the most diluted case considered here, the relative decrease in dipolar field is approximately 20% and reductions to more than 100% for the highest concentration considered are found. For the case of sample R1 (for $c = 0.3$) this reduction is estimated to be about 100 %. This gives support to the interpretation of the changes in the effective energy barriers observed experimentally for sample R1 as compared to R2 when increasing the cooling field, since changes in the dipolar fields acting on the particles are directly related to the corresponding modification of the energy barriers due to the anisotropy itself. As an estimate of the increase in the energy barriers induced by dipolar interactions, we have calculated the increase in the mean energy barrier by introducing the mean dipolar field obtained in the simulations in the expression for a nanoparticle with easy axis aligned along the field direction $E_b = E_b^0 (1 + H_{dip}/H_{ani})^2$, where $E_b^0/k_B = KV_m/k_B = 175$ K is the mean energy barrier for non-interacting particles of sample R2. Plugging the estimated values of $H_{dip} = 114$ Oe (for $c = 0.3$) and $H_{ani} = 1240$ Oe gives $E_b/k_B = 209$ K which is in close agreement with the peak position in Fig. 5 for sample R1 ($T_{max} = 217$ K). Consequently, those results unambiguously demonstrate that the observed differences between the distributions of energy barriers for samples R1 and R2 are essentially due to the effect of dipolar interactions present in R1.

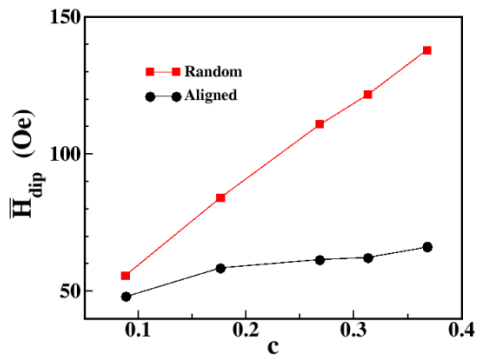

**Figure 7.** Concentration dependence of the mean value of the dipolar field modulus generated by 4096 particles placed at random in a 3D box of size $aL$. Only the fields on the central part of the simulated box have been taken into account. Black circles correspond to a configuration with magnetic moments aligned along the field direction while red squares correspond to randomly oriented magnetic moments.

## Conclusions

We have shown that the average dipolar field acting on each individual particle in an ensemble of interacting particles is strongly reduced when the particles' magnetizations are collinearly aligned, this reduction being more significant as the volume concentration of the magnetic cores increases. As a consequence, we propose a general method to quantify the effect of dipolar interactions on the effective distribution of energy barriers for magnetization reversal, by comparing the master curve for the magnetization relaxation obtained after field cooling the sample at low field (initial state with random orientation of the particles' magnetizations) to that corresponding to intermediate fields where the particles' magnetizations tend to be aligned along the field direction in the initial state. We have also obtained the intrinsic distribution of the energy barriers of anisotropy for the non-interacting case by studying the relaxation of the magnetization for a reference sample where the same magnetic particles were individually coated with a thick silica shell in order to make dipolar interactions negligible. Interestingly, numerical simulations account for the observed energy shift between the distribution of energy barriers corresponding to the interacting and non-interacting cases, thus supporting the proposed method for the quantification of dipolar interactions.

**Supporting Information available**. Particle size distributions of the samples, X-ray patterns of the samples. This material is available free of charge via the Internet at http://pubs.acs.org.

**Corresponding Author**
*Corresponding author. Email: cmoya@ffn.ub.es

**Author Contributions**
The manuscript was written through contributions of all authors. All authors have given approval to the final version of the manuscript.

**Acknowledgment**
This work was supported by Spanish MINECO (MAT2012-33037) and Catalan DURSI (2014SGR220).